          [2001/04/25 1.1 (PWD)]
\documentclass[a4paper,twocolumn]{esapub} 
\usepackage{natbib,graphicx}

\title{INTEGRAL observation of Cyg X-1 in an intermediate state}
\author[1,2]{J. Malzac}
\author[3]{P.O.~Petrucci}
\author[1]{E.~Jourdain}
\author[4]{P.~Sizun}
\author[4]{M.~Cadolle}
\author[5]{G.~Pooley}
\author[3]{C.~Cabanac}
\author[4,6]{S.~Chaty}
\author[7]{T.~Belloni}
\author[4,8]{J.~Rodriguez}
\author[1]{J.P.~Roques}
\author[4]{P.~Durouchoux}
\author[4]{A.~Goldwurm}
\author[4]{P.~Laurent}
\affil[1]{Centre d'Etude Spatiale des Rayonnements, 31028 Toulouse, France}
\affil[2]{Institute of Astronomy, Madingley road, CB3 0HA, Cambridge, UK}
\affil[3]{Laboratoire d'Astrophysique Observatoire de Grenoble, BP 53
F-38041 GRENOBLE C\'edex 9, France}
\affil[4]{Service d'Astrophysique, DSM/DAPNIA/SAp, CEA-Saclay, Bat. 709,
L'Orme des Merisiers, F-91 191 Gif-sur-Yvette, Cedex, France  }
\affil[5]{Cavendish Laboratory, University of Cambridge, Madingley
Road, Cambridge CB3 0HE, UK }
\affil[6] {Universit\'e Paris 7 Denis-Diderot 2 place Jussieu 75251
Paris Cedex 05, France}
\affil[7] {INAF - Osservatorio Astronomico di Brera, via E. Bianchi 46, 23807 Merate, Italy} 
\affil[8]{INTEGRAL Science Data Center, Chemin d'\'Ecogia 16, 1290 Versoix, Switzerland}

\begin{document}


\maketitle

\begin{abstract}
We present preliminary results of an observation of Cygnus X-1 with
INTEGRAL performed on June 7-11 2003.
 Both spectral and variability properties of
the source indicate that Cygnus X-1 was in an intermediate state.
As expected during state transitions, we
find an anticorrelation between the 3-10 keV and the 15 GHz radio fluxes
and a strong correlation between the 3-200 keV hardness and radio flux.
\end{abstract}

\section{Introduction}

\begin{figure}[!t]
\centering
\includegraphics[width=\linewidth]{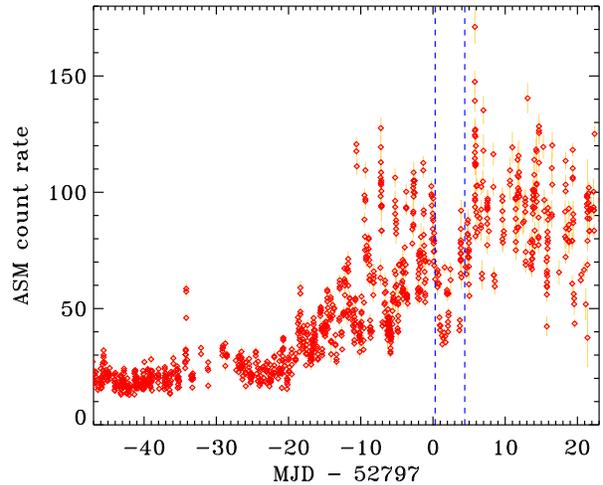}
\caption{RXTE/ASM light curve of Cygnus X-1. The time of the INTEGRAL
observation  is delimited by the vertical dashed lines \label{fig:asm}}
\end{figure}

Cygnus X-1 is the prototype of black hole candidates. 
Since its discovery in 1964 (Bowyer et al. 1965), it has been intensively observed by 
all the high energy instruments, from soft X-rays to $\gamma$-rays. 
It is a persistent source most often observed
 in the so-called low/hard state, characterised 
by a relatively low flux in the soft X-rays ($\sim$ 1 keV) and a
high flux in the hard X-rays ($\sim 100$ keV).
In the hard state, the high energy spectrum can be roughly described
 by a power-law 
with spectral index $\Gamma$ varying in the range 1.4-2.2, 
and a nearly exponential cut-off at a characteristic energy 
$E_{\rm c}$ of a few hundred keV (see e.g. Gierlinski et al. 1997).
Occasionally, the source switches to the high soft state. The
high energy power-law is then much softer ($\Gamma > 2.4$) and the
bolometric luminosity is dominated by 
a thermal component peaking at a few keV.
Finally, there are also intermediate states
in which the source exhibit a
relatively soft hard X-ray spectrum ($\Gamma \sim 2.1-2.3$) and a
moderately strong soft thermal component (Belloni et al. 1996; Mendez \& van der Klis 1997). The intermediate state often,
but not always, appears when the source is about to switch from one
state to the other.
When it is not associated with a
state transition, it is interpreted as a 
`failed state transition'.
Until 1998 the source used to spend nearly 90 \% of its time in the
hard state. In the recent years however 
      there have been more intermediate states and soft states 
(see Zdziarski et al. 2002,  Pottschmidt et al. 2003, Gleissner et al. 2004).
Cygnus X-1 represents a prime target for the INTEGRAL mission (Winkler et
al. 2003) launched in 2002 October, which instruments cover an unprecedented
spectral coverage at high energy, ranging from 3 keV to several MeV.
Cygnus X-1 was extensively observed (600 ks) during the calibration
phase of the mission (Bouchet et al. 2003, Pottschmidt et al. 2003,
Bazzano et al. 2003). At that time, the source presented all the
characteristics of the hard state.
In this paper we report preliminary results of the first
observation of Cygnus X-1 in the open time programme. 
This  300 ks  observation was performed on 2003 June 7-11 (rev 79/80)
with a 5x5 dithering pattern (the effective exposure time was 275 ks
for JEM-X2, 292 ks for IBIS, and 296 ks for SPI).
At this epoch, the RXTE All Sky Monitor count rate of Cyg X-1 was
larger than in typical hard states by up to a factor of 4, and the
light curve of Cyg X-1 shows strong X-ray activity characteristic of
state (or failed state) transitions 
(see Fig.~\ref{fig:asm}).  We also combine the INTEGRAL data with
the results of coordinated radio observations (15 GHz) performed
with the Ryle telescope.     

\begin{figure}[!t]
\centering
\includegraphics[width=\linewidth]{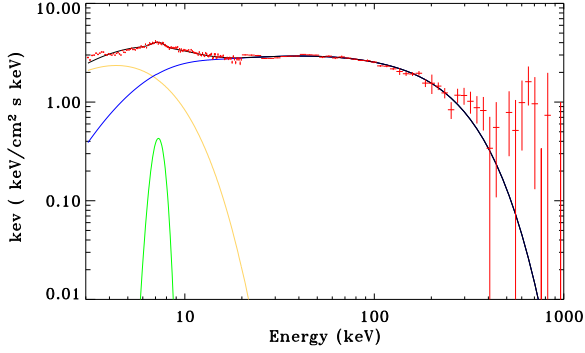}
\caption{Joint JEM-X/SPI spectra of Cygnus X-1 averaged over revolution
79. \label{fig:spec79} }
\end{figure}
\begin{figure}[!ht]
\centering
\includegraphics[width=\linewidth]{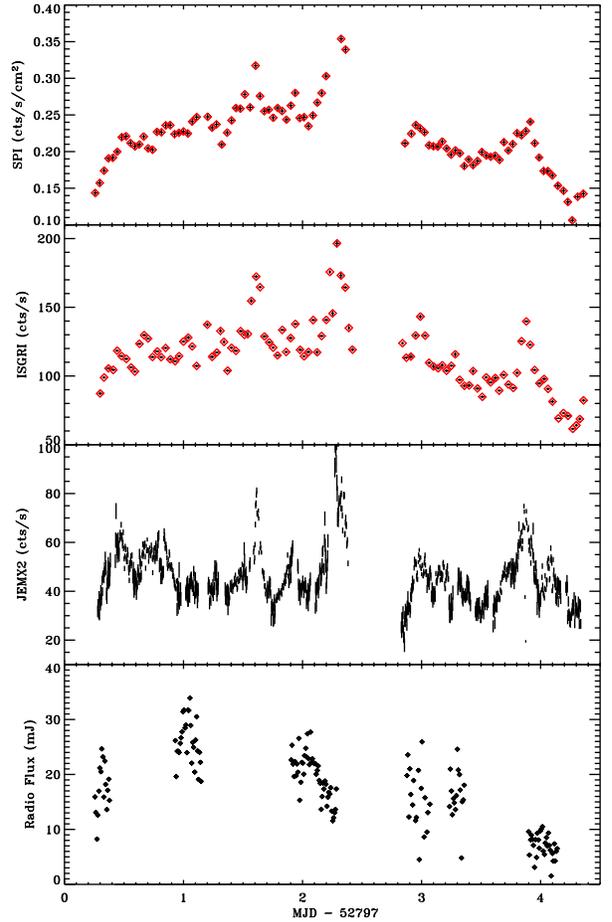}
\caption{Light curve of Cygnus X-1 as measured by SPI (20-50 keV),
ISGRI (15-400 keV), JEM-X (3-15 keV)
and the Ryle telescope (15 GHz).\label{fig:lc} }
\end{figure}
\begin{figure}[t]
\centering
\includegraphics[width=\linewidth]{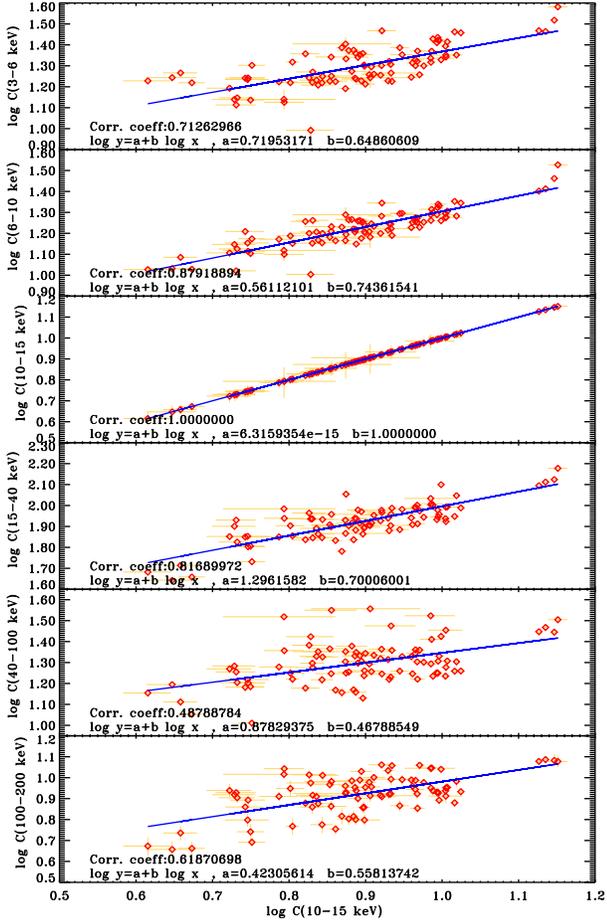}
\caption{Sample flux vs flux correlations (30~min pointings) for JEM-X
($<$15 keV) and ISGRI ($>$ 15 keV), $C$ denote count rate in counts/s.
 \label{fig:flfl}}
\end{figure}

\begin{figure}[t]
\centering
\includegraphics[width=\linewidth]{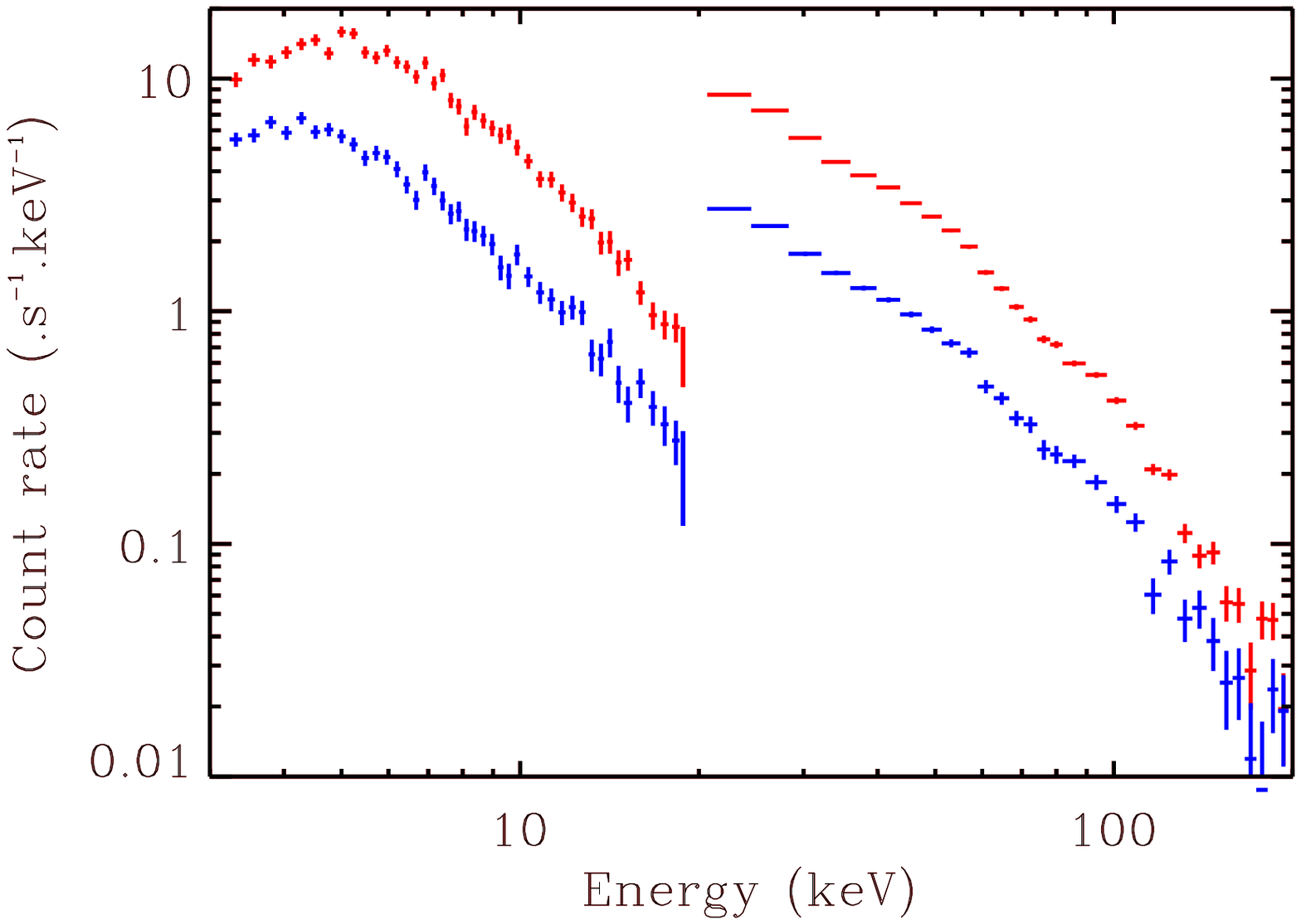}
\includegraphics[width=\linewidth]{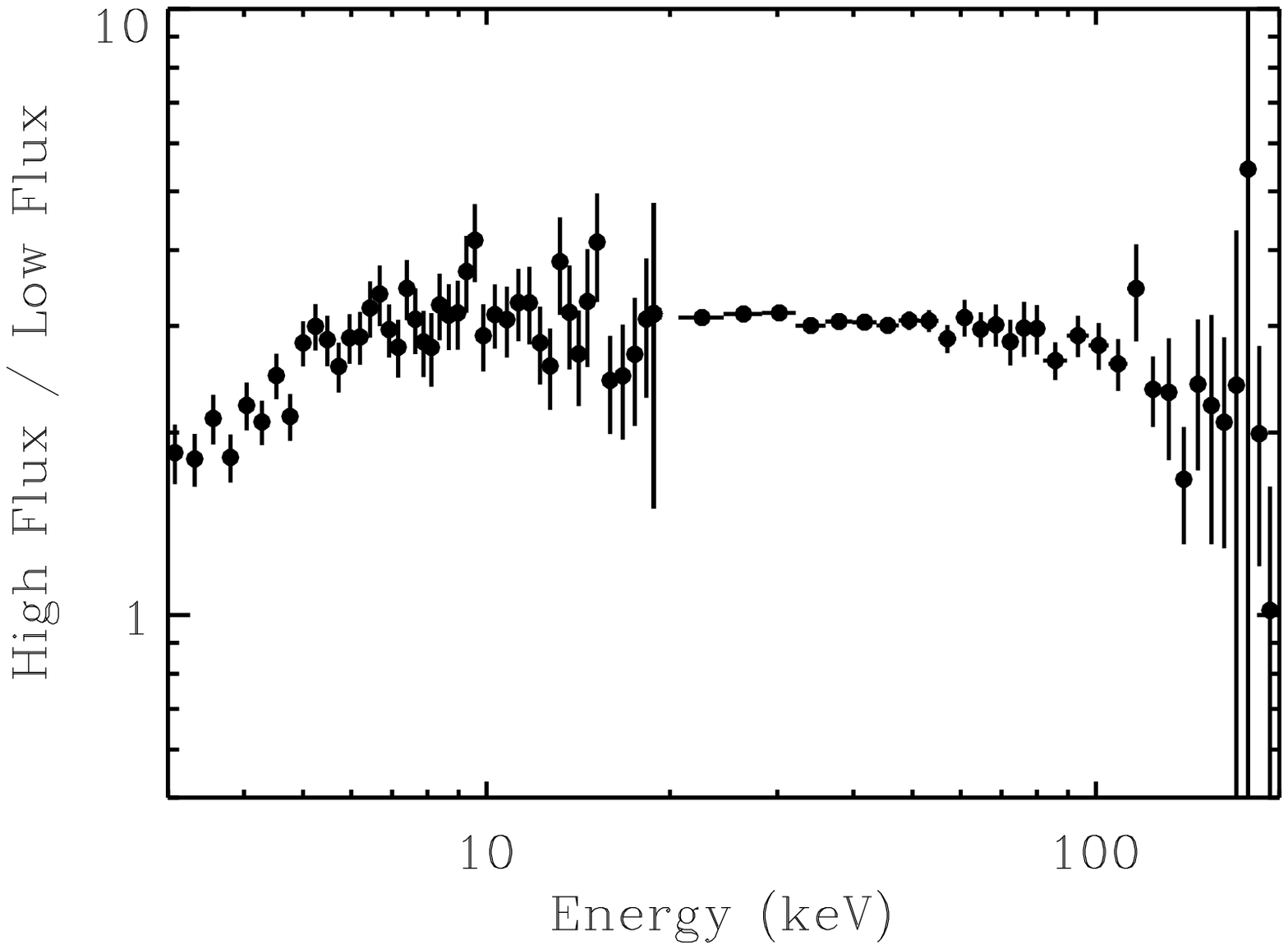}
\caption{
30~min exposure JEM-X/ISGRI spectra  at highest and lowest observed ISGRI flux (top)
and their ratio (bottom).  
 \label{fig:specpos}}
\end{figure}

\begin{figure}[!ht]
\centering
\includegraphics[width=\linewidth]{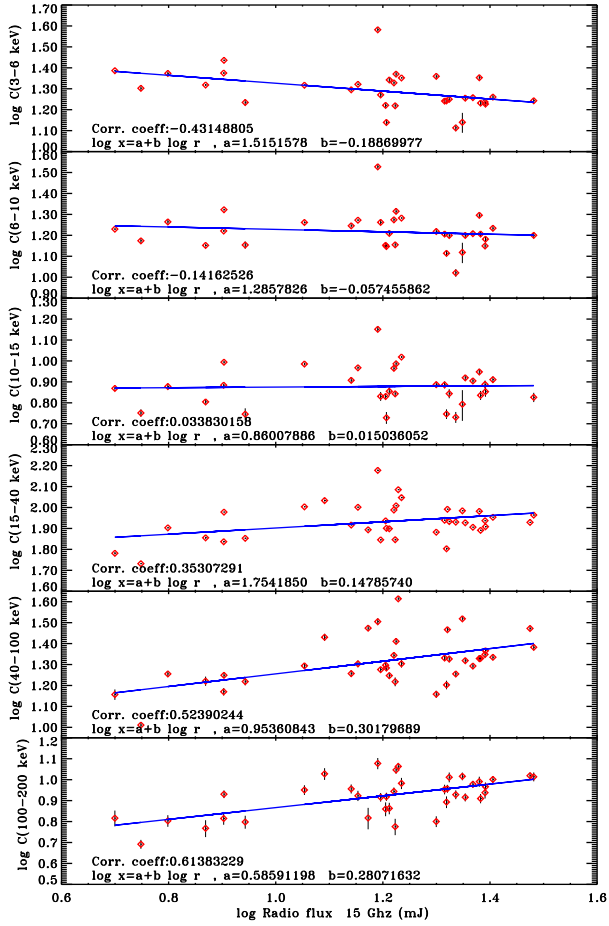}
\caption{Count rate in different JEM-X ($<$15 keV) and ISGRI ($>$ 15 keV)
bands versus 15 GHz radio flux density \label{fig:radiocor}}
\end{figure}
\begin{figure}[!ht]
\centering
\includegraphics[width=\linewidth]{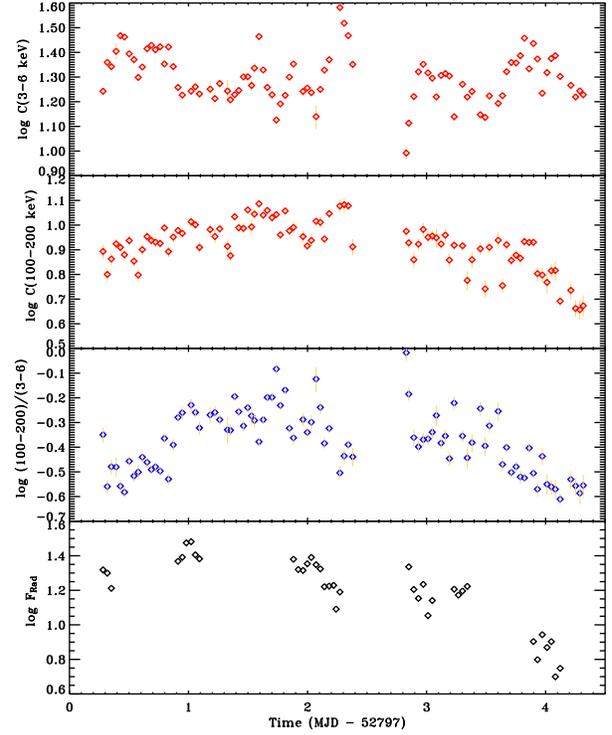}
\caption{Time evolution of (from top to bottom) the 3-6 keV JEM-X count
rate, 100-200 keV ISGRI count rate, (100-200 keV)/(3-6 keV) hardness
ratio, and radio flux density (all averaged over 30 min). \label{fig:lcband}}
\end{figure}
\begin{figure}[!ht]
\centering
\includegraphics[width=\linewidth]{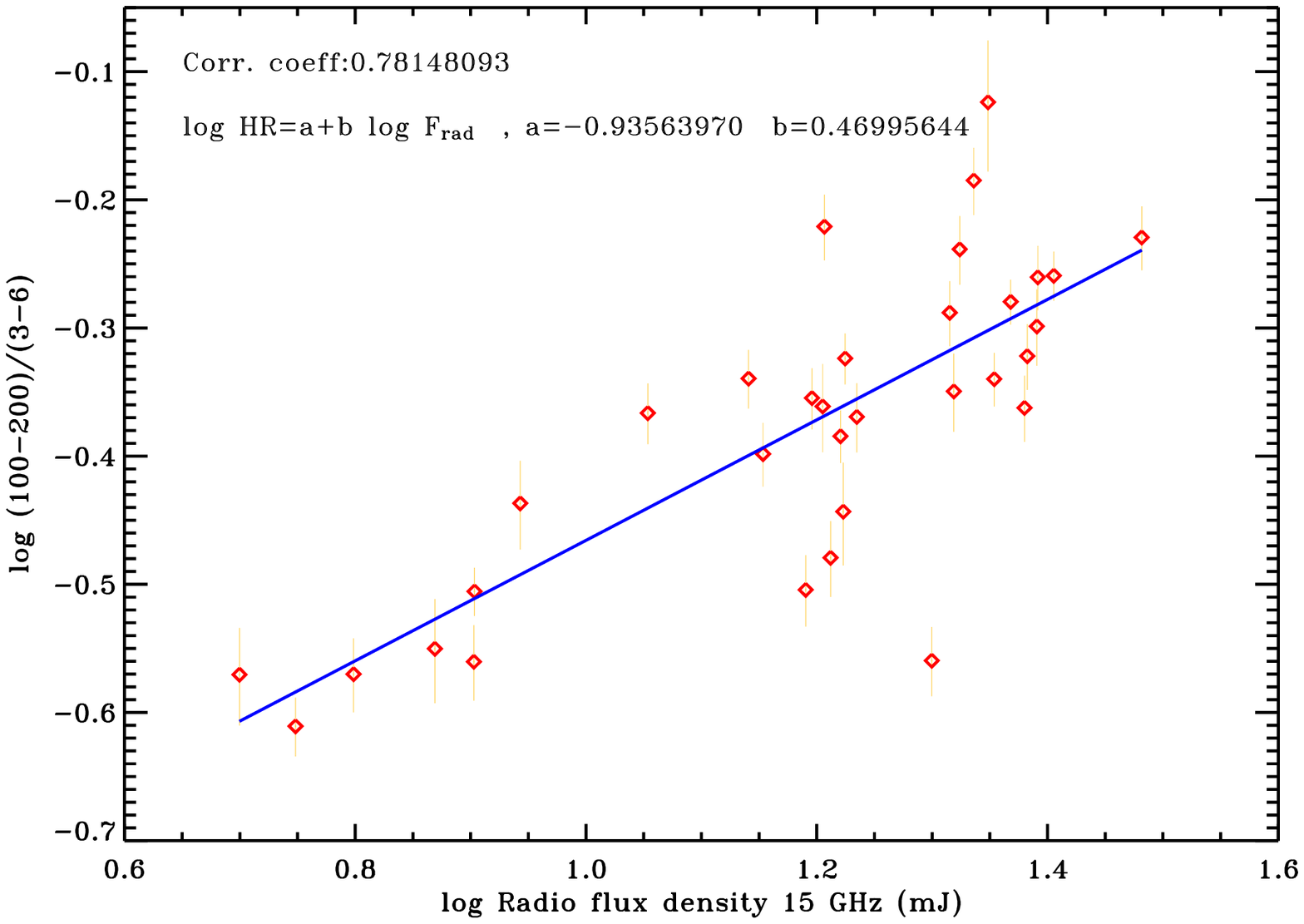}
\caption{ Correlation between radio flux density and (100-200
keV)/(3-6 keV) hardness \label{fig:radiohr}}
\end{figure}
\section{Spectrum and variability}
The data were processed using the OSA-3 software. Above 20 keV, given the current
calibration of ISGRI, we restrict our spectral modelling  to the SPI
data. In addition, we use JEM-X in the 3-20 keV but warn that
significant calibration problems are reported below 10 keV. 
The broad band JEM-X/SPI spectrum cannot be represented by a single power-law.
 In addition, a strong soft component affecting the whole 3-15 keV 
band is required. On the other hand, at high energies, the 
SPI data provide a highly significant evidence
for a cut-off (or at least, a break) at about 100 keV.
In order to estimate the intrinsic spectral slope, we fitted the
joint JEM-X/SPI spectrum in the 8-50 keV range, where the effects of the soft
component and high energy break, are not too strong.
The best fit spectral index was $\Gamma$=2.14$\pm$0.1. 
The spectrum is thus rather soft with respect to typical hard state spectra ($\Gamma$$\sim$1.7), but harder than  the soft state spectra ($\Gamma$$\sim$2.5).
The modelling of the broad band (3 keV-1 MeV) combined JEM-X/SPI 
spectrum definitely requires a multi-component model.  
As a preliminary attempt, we used a comptonised spectrum (COMPTT,
$kT_e$=60 keV, $\tau$=0.68) plus a disk blackbody   ($kT_{\rm in}$=1.7
keV) and a Gaussian line ($E$=7.1 keV). The average deconvolved
spectrum and model are shown in Fig.~\ref{fig:spec79}. 
We also attempted to include a neutral reflection bump component but 
the fit resulted in a very low reflection amplitude (R$<$2
10$^{-2}$). The apparent absence of
reflection hump together with a strong iron line in the JEM-X spectrum
is really puzzling.  This could imply that the
reflecting medium is strongly ionised and/or relativistically
smeared and we plan to investigate such models.
 However, calibration problems in the JEM-X
spectrum and/or the lowest energy range  of SPI is also a strong possibility.
We further note that the value of the fit parameters are only
indicative since, as already mentioned, large uncertainties 
in the calibration remain in the
present version of the software. Moreover, intra-observation spectral
variability could also affect the shape of the averaged spectrum and
the physical parameters derived from the fitting procedure.
Despite these uncertainties, we can conclude that together with the 
behaviour of the ASM light curves, 
the relative softness of the spectrum as well as the presence of the 
strong soft component in the JEM-X band suggest that the source was 
in an  intermediate state.
 
The light curves provided by JEM-X, SPI, IBIS and the Ryle telescope
are shown in Fig.~\ref{fig:lc}. We note that the JEM-X and ISGRI light
curves are affected by the dithering mode. For instance depending on the
offset angle the ISGRI flux of a source can vary by more than
 20 \%, especially below 40 keV. 
On time scales of hours, the variability appears to be dominated by changes in the overall luminosity 
with only weak  spectral variability. In the energy range considered
(3-200 keV), the 30 min average fluxes in different
 bands are all correlated with each other (see Fig.~\ref{fig:flfl}).
There is however significant evidence for
spectral changes  
when the overall luminosity increases. Fig.~\ref{fig:specpos} compares
the spectra obtained for the minimum and maximum luminosities in our
observation. 
This figure shows that the higher luminosity spectrum is slightly 
softer above 100 keV and much harder 
below 7 keV.
The latter, if not caused by
calibration problems, could be due to a
larger temperature of the thermal disc component in the higher flux
pointing. On the other hand, in the 7-100 keV band, there is no hint
for any spectral variability despite the overall luminosity differing
by a factor of about 3. 
\section{Radio/High energy correlation}
The radio flux tends to be anti-correlated with the X-ray flux (3-15
keV)  and correlated with the soft gamma-rays ($>$15 keV). This
dependence of the INTEGRAL/radio flux correlation (shown in
Fig.~\ref{fig:radiocor}) suggests that the fluctuations of the 
radio luminosity is associated with a pivoting of the high energy
spectrum, with the pivot point located around 10-15 keV.
As a consequence, there is a strong correlation between the hardness
of the high energy spectrum and the radio flux (see
Fig.~\ref{fig:lcband} and Fig.~\ref{fig:radiohr}). 
Therefore, on time scales of hours, the  radio jet
activity is correlated with hardening of the high energy spectrum  
rather  than high energy luminosity.
This result strongly differs from what is usually reported in the hard
state of Cygnus X-1 and other sources. Indeed, the radio flux is then
 positively correlated with the soft X-ray emission (3 - 25 keV,
Corbel et al.  2000,
2003; Gallo, Fender, Pooley 2003). 
Nevertheless, as mentioned above,  the variability  and spectrum of the
source suggest that during our observation, Cygnus X-1 was not in a
typical hard state but 
in an intermediate state. 
Actually, the transition from hard to soft state is known to be 
associated with a quenching of the radio emission (Corbel et al. 2000;
Gallo, Fender, Pooley 2003). As the transition to the soft state
also corresponds to a strong softening of the spectrum, this is
consistent with the correlation between hardness and radio flux: 
when, during the observation, the source gets closer to the soft state
the spectrum softens and simultaneously the radio flux decreases.
We note that a recent analysis of Ryle and RXTE data of Cyg X-1 (Gleissner et
 al., 2004, A\&A, submitted) interestingly shows the same
correlation tendencies during failed state transitions (Ryle/PCA:
moderate anti-correlation, Ryle/HEXTE: correlation) as reported
here, albeit on timescales from weeks to years.


\begin{thebibliography}{}

\bibitem[Bazzano et al.(2003)]{2003A&A...411L.389B} 
Bazzano, A., et al.\ 2003, A\&A, 411, L389 

\bibitem[Belloni et al.(1996)]{1996ApJ...472L.107B}
 Belloni, T., Mendez, M., van der Klis, M., et~al., 1996, ApJl, 472, L107 

\bibitem[Bouchet et al.(2003)]{2003A&A...411L.377B}
 Bouchet L., et al.\ 2003, A\&A, 411, L377
 
\bibitem[bowyer et al. (1965)]{b1965}
Bowyer, S., Byram, E.T., Chubb, T.A., Friedman, M., 1965, Sci, 147, 394

\bibitem[Corbel et al.(2000)]{2000A&A...359..251C}
 Corbel, S.,
et~al.,
\ 2000, A\&A, 359, 251 

\bibitem[Corbel et al.(2003)]{2003A&A...400.1007C} Corbel, S., Nowak, 
M.~A., Fender, R.~P., Tzioumis, A.~K., \& Markoff, S.\ 2003, A\&A, 400, 
1007 



\bibitem[Gallo, Fender, \& Pooley(2003)]{2003MNRAS.344...60G} 
Gallo, E., Fender, R.~P., \& Pooley, G.~G.\ 2003, MNRAS, 344, 60 


\bibitem[Gierli{\' n}ski et al.(1999)]{1999MNRAS.309..496G} 
Gierli{\' n}ski, M., 
 et~al.,
 MNRAS, 309, 496 

\bibitem[Gierlinski et al.(1997)]{1997MNRAS.288..958G} 
Gierlinski, M., et~al.,
\ 1997, MNRAS, 288, 958 


\bibitem[Gleissner et al.(2004)]{2004A&A...414.1091G}
 Gleissner, T., 
et~al.,
 A\&A, 414, 1091 



\bibitem[Mendez \& van der Klis(1997)]{1997ApJ...479..926M}
 Mendez, M.~\& van der Klis, M.\ 1997, ApJ, 479, 926 


\bibitem[Pottschmidt et al.(2003)]{2003A&A...411L.383P}
 Pottschmidt, K., et al.\ 2003, A\&A, 411, L383 


\bibitem[Winkler et al.(2003)]{2003A&A...411L...1W}
 Winkler, C., et al.\ 2003, A\&A, 411, L1 


\bibitem[Zdziarski, Poutanen, Paciesas, \& Wen(2002)]{2002ApJ...578..357Z} 
Zdziarski, A.~A., Poutanen, J., Paciesas, W.~S., \& Wen, L.\ 2002, ApJ, 
578, 357 


\end{thebibliography}
\end{document}